\documentclass [12pt] {article}

\usepackage{graphicx}
\DeclareGraphicsExtensions{.pdf,.png,.jpg}

\begin {document}

\noindent {\large \bf Cartesian nature of  equal active and passive energies}

\bigskip \noindent{I.E. Bulyzhenkov}
\smallskip

\noindent{{\small Moscow Institute of Physics \& Technology and 
Lebedev Physics Institute RAS, Moscow, Russia, bulyzhenkov.ie@mipt.ru}}\\
\bigskip

 {   { \small Abstract. The Cartesian material space approach to  Maxwell's equations reveals the analytical solution of the continuous radial density for the extended elementary charge.  
 Radial charges and their Coulomb fields carry equal passive and active imaginary energies. The Einstein mass-energy formula can be updated for electricity in complex numbers.  The Principle of Equivalence works universally  for active/passive masses and for active/passive electric charges. Maxwell's equations for complex continuous charges become 4-vector identities for locally paired flows of energies that correspond to the Chinese 
  yin-yang approach to the nature.  
 Cartesian relativistic physics has its own nondual analog of the Einstein Equation, which leads to geodesic equations for the motion of  Ricci material densities and to vortex states of the elementary mass-extension.    Non-empty space physics modifies the Navier-Stokes equation by kinematic \lq living forces\rq{}. This inertial feedback enables a conceptual choice between Newtonian and Cartesian universes (with localized or extended elementary masses, respectively) because of new predictions of pressure and temperature gradients across laboratory flows of liquids and gases.
}}



\bigskip 
\bigskip 

Cartesian physics \cite {3,Car} of material space is currently described only in qualitative terms and without quantitative predictions of specific phenomena for  the theory verification procedure. This physics refers to Descartes\rq{s} natural philosophy \cite {1,2}
which is still \lq  {\it The Terra Incognita}\rq{} for pragmatic supporters of the dual Newton world with empty space between localized mechanical bodies. 
  The concept of void space regions without matter was unclear to Descartes, who after Aristotle also maintained the extension of matter or \lq matter-extension\rq{} as the continuous material plenum. The Cartesian vortex mechanics for all types of observable spatial displacement in this material plenum  was published in 1629. Later Newton's  successful dynamics of point-like masses shook the need in  sophisticated matter-extensions with vortex states. Nowadays, Newton\rq{s} mechanics of localized masses and continuous gravitation fields in empty space form the dual core of contemporary space theories in the Solar system and beyond. Not Cartesian physics, but Newtonian empty space modeling became mandatory for 1916 Einstein's gravitation in the weak field limit. No one looked at mechanical theories if they did not fit  Newton\rq{s} mass transport at low speeds.

Western science invented the motto  \lq Nullius in verba\rq{} in a line of mandatory experimental evidence. This pragmatic approach to reality turned out very promising for initial technical  progress that took place since the Middle Ages. Western researchers and engineers tend to be independent from any religion and  from the influence of any philosophical teachings, including Cartesianism of material space and  Chinese reading of immeasurable  reality. Recall that the ancient Chinese science has recognized the  coexistence of  corporeal and incorporeal (immeasurable) worlds that contradict the  Western science scope. However, many eastern findings, like the traditional Chinese medicine,  are still challenging western researches.

The 1961 Eightfold way of M.Gell-Mann and other formal \lq coincidences\rq{} between modern physics and  ancient Eastern dogmas have been mentioned by many authors, for example by F.Capra \cite{Cap}. Contemporary physicists merely superficially trace the roots of the Standard Model in ancient texts and symbols like the eight trigrams. In conventional scientific publications, the authors avoid referring to Eastern teachings. The latter originated, in particular, from the classic \lq Book of Changes\rq{} (\lq I Ching\rq{}), a fundamental work in the Chinese history and culture. In general, it would be wrong to say that many leading researchers were inspired by the philosophical treasury, whether it be East or West.

By rejecting invisible material phenomena beyond the limits of laboratory measurements, it is impossible to fully understand, for instance, Cartesian natural philosophy, the yin-yang Chinese dialectics, and the Russian Cosmism \cite{Rus} principles (\lq\lq{}people are citizens of the whole Universe rather than the Earth\rq\rq{}). Beginning with Plato, the Ancient Greeks considered space to be a continuous material plenum, filled everywhere by \lq the most translucent kind (Aether)\rq{}. Material bodies cannot be replaced, according to Aristotle\rq{s} logical proof, to completely empty, intangible place outside the continuous space-plenum. Descartes followed this  logical avenue and eventually offered his vortex mechanics for the spatial transport of variable energy charges. Should we neglect Cartesian physics of non-empty space due to \lq laboratory evidence\rq{} in favor of the Newtonian empty space model?  The point mass notion has to be conceptually criticized  by the relativistic carrier of variable internal energy, which exhibits the  proper time dilatation (frequency red shift) in the case of one moving particle or the kinematic cooling in the case of a moving thermodynamical body. Due to low speed changes of  bodies' internal energies, Einstein's Theory of  Relativity better corresponds the  Cartesian world of viriable energy charges in the low speed limit than the Newtonian world of constant gravitational/inertial charges. Below we plan to describe  this statement quantitatively and to reject Newtonian references from the self-contained metric description of non-empty inertial spaces. 

Dual classical physics for spatially separated matter and fields  is traditionally assigned to the macroscopic world, while nondual material densities of  quantum fields physicists  tend assign to only the microscopic world. But the unique physical reality is either dual or nondual regardless of the spatial scaling and mathematical  formalisms in available theories. Similarly, the matter is either local or non-local regardless of suitable approaches to describe it.   The celebrated Einstein-Podolsky-Rosen paradox initiated the long-term discussion and many tests of the material world nonlocality. Now nonlocality of matter is reasonably considered beyond quantum physics \cite {PopE}. The world holism \cite {Smu} and the global direct overlap of all material elements motivates our attempts to develop non-Newtonian mechanics for mass densities of the extended particle. The Newton empty space model vs the Cartesian material space of  overlapping extended vortexes is illustrated in Figure 1. There is no in reality a mesoscopic scale for a possible transitions from nondual microscopic physics to dual macroscopic model of Newton. Material states should initially be considered  in nondual field terms on micro, macro, and mega scales in qualitative, if not quantitative, approaches to any energy flows within the Universe.

\begin{figure*}
\centering
\includegraphics[width=6cm,clip]{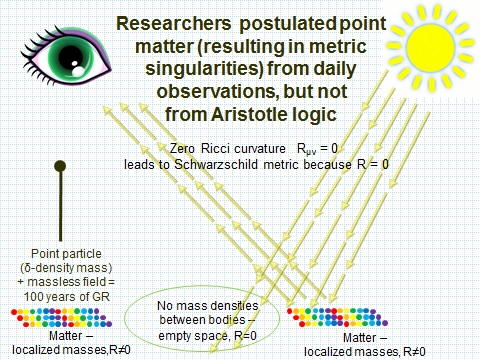}
\includegraphics[width=6cm,clip]{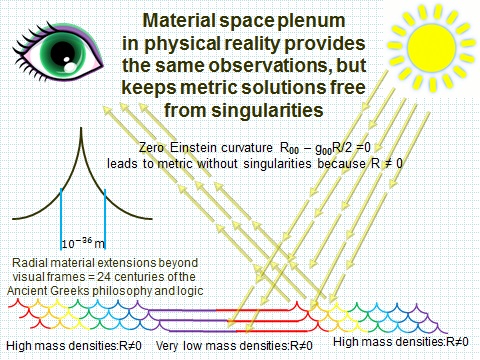}
\vspace*{0.2cm}       
\caption{Newtonian particles and empty space vs Cartesian nondual plenum}
\label{fig-1}       
\end{figure*}

 The purpose of this paper is to trace the balanced energy origin of extended charges in Maxwell's electrodynamics and Einstein's metric gravitation, to study strong field energy conservation and compensation for overlapping material elements in terms of complex densities, to propose nondual field analogues of the Maxwell equations and the Einstein equation, and to predict inertial space phenomena for conceptual tests of Cartesian physics in the laboratory. 
Based on propositions of Aristotle, Descartes, Mie \cite{Mie} and Einstein \cite {Ein}, we shall try to study how to redesign classical electrodynamics and gravitation in  pure field terms (material fields without localized particles) in order to construct a bridge between the Western science pragmatism and the eastern untestable approach to the invisible reality. Our initial idea is that \lq Nullius in verba\rq{} physics may coherently replace the supposed  empty space (between the localized charges) with the Aristotle space plenum of charged material fields. This may assist to give up the unnecessary concept of Newton's point-like particles and to modify the old   building of classical gravitation and electrodynamics by strong field solutions without energy divergence.   

Nonempty space with a  continuous charge density  and a density of the continuous mass will have the  energy meaning in our computations. All material densities will obey the universal Yin-Yang Principle for local pairing of complementary (passive and active) energy flows. This Chinese principle was rediscovered by Einstein as the Principle of Equivalence for gravitational (passive) and inertial (active) masses of  mechanical (corporeal) bodies.  Below we  prove analytically that the Yin-Yang Principle can be applied to gravitational / mechanical field densities of continuous masses. The same is true for local pairing of active / passive energy densities in Maxwell's electrodynamics.   The yin-yang dialectic suggests in such a way to unify as the charge with its field, as well as electricity with gravity. For this, we used complex numbers, while four known interactions may be unified similarly in algebra of quaternions. 

The Einstein Equation in the dual (field + matter) Newtonian paradigm has been known since 1915. The similar tensor equation in the Cartesian paradigm of the united space-matter operates with a nondual field continuum of extended mass-energies. This continuum is filled by radial vortexes with chaotic auto-rotations of elementary metric densities. Such permanent metric motions within the elementary Cartesian vortex  reveal the kinematic origin of the rest-mass  energy  $mc^2$. This active, kinetic energy may be called the internal relativistic heat. The positive (kinetic, active) energy is always accompanied by equal, but negative (potential, passive) self-energy according to the Yin-Yang Principle.  Therefore, the mass creation mechanism from \lq the void nothing\rq{} can be quantitatively discussed in Cartesian physics and in Chinese dialectics. There is no elementary inertial  mass without vortex auto-rotations of metric space. And circular material densities  with positive kinetic energy generate self-gravitation with negative (passive) elementary energies of Cartesian vortex states that results in the global compensation of active and passive energies in the non-empty space Universe.

We will return to the Einstein idea that the geodesic equations of motion should follow from the strong-filed gravitation equation. This was difficult to prove from iterations in the dual post-Newtonian physics. But Cartesian geodesic relations for Ricci material densities  can be derived exactly from the nondual field analogue of the Einstein Equation.   In order to demonstrate some practical benefits of non-empty space physics, we will modify the Navier-Stokes equation in terms of Ricci energy flows with the inertial feedback of \lq living\rq{} forces, $-\mu \partial_i V^2/2$. This dynamical pressure averts asymptotically divergent energy flows and can initiate conceptual tests of Newton vs Cartesian mechanics in the laboratory.

This paper is not a polemical overview of the Descartes philosophical treasury or the known neo-Cartesian approaches to the nature. We focused on the analytical energy balances for the introduced radial charge instead of the point particle in order to reinforce the non-empty space idea for Maxwell's electrodynamics,  for     
Einstein's special and general theories of relativity, and for thermodynamics of condensed matter. The following mathematical exercises may be useful also for those researchers of living matter, who would like to look at the world holism and nonlocality beyond the formal procedures of quantum mechanics.

   \bigskip 
  \noindent {\bf Coulomb fields  within continuous  charges} 
   \smallskip

The predominant majority  of people believe that electric charges and inertial masses are located within the visible frame of  macroscopic bodies. Such bodies can be divided into smaller parts in accordance with everyday  observations. Thus, the smallest part of substance was  introduced  as a so-called corpuscle (the point particle in the indivisible limit) that is a carrier of the elementary mass and charge. Classical fields between localized particles in  \lq Nullius in verba\rq{} physics are considered massless and chargeless. Nevertheless, the experiment is only a criterion of truth, but not the truth itself. The materialistic pragmatism of highly educated and well-equipped researchers of the supposed 
dual  \lq reality\rq{} (particle and field are different notions) was rejected by many philosophers. Indeed, by dividing an infinitely extended field-energy object into parts one should again obtain only infinite objects rather than corpuscles of limited size. 

In contrast to available observations,   nonempty space of continuous material flows has been recognized by philosophers not only in the Ancient East and in  Ancient Greece, but also by many  
contemporary  thinkers in the West (Mie, Einstein, Infeld) and in Russia (Fedorov, Umov, Vernadsky, Roerich {\it et al.}\cite{Rus}). \lq\lq{}A coherent field theory requires that all elements be continuous... And from this requirement arises the fact that the material particle has no place as a basic concept in a field theory. Thus, even apart from the fact that it does not include gravitation, Maxwell's theory cannot be considered as a complete theory\rq\rq{}as was stated by Einstein and Infeld  \cite{Ein} in 1938. Indeed, the postulated  point-particle paradigm results in Coulomb divergence, which terminates Classical Electrodynamics as a self-consistent theory.  A point source in the Maxwell-Lorentz Equations may be considered as \lq\lq{}an attempt which we have called intellectually unsatisfying\rq\rq{} according to De Broglie \cite{Bro}. Einstein also criticized his 1915 field equation for the point gravitational source: \lq\lq{}it resembles a building with one wing built of resplendent marble and the other built of cheap wood\rq\rq{} (translation \cite{Ton}). 
 The Dirac delta-operator for point material densities in  emptiness seem  to have pushed Western physicists far away from the physical reality and Eastern approaches to superimposed energy flows. 

Continuously distributed elementary charge was reasonably inferred by Mie in order to derive properties of nonlocal charges from properties of their EM fields and four-potentials. Regretfully, in 1912-1913
the \lq{}Theory der Matter\rq{} \cite{Mie} had not found the exact (logarithmic) potential of strong fields within nonlocal classical matter.  Therefore, the  promising non-empty space concept had not been proven before the mathematical era of quantum non-locality began. Formal probabilities of  the delta-operator \lq{}dice\rq{} in the Newtonian empty space arena postponed  search of rigorous analytical solutions for Mie\rq{}s nonlocal world of continuous particles.

The classical  field equations keep the analytical solution option for a distributed elementary particle. By following Mie, we assume that it is possible to relate locally  electric, ${\vec d}(x)$, and magnetic, ${\vec b}(x)$, field intensities in the Maxwell-Lorentz equations \cite{L}, 
\begin {eqnarray}
  \cases { div\ {\vec {d}}({{\bf x}},t) = 4\pi\rho({{\bf x}},t) \cr
div\ {\vec b}({{\bf x}},t) = 0 \cr
c\ curl\  {\vec b}({{\bf x}},t)  = 4\pi\rho ({{\bf x}},t){\vec v} + \partial_t  {\vec {d}}({{\bf x}},t)\cr
c\ curl\ {\vec {d}} ({{\bf x}},t)  = - \partial_t {\vec b}({{\bf x}},t),\cr}
\end {eqnarray}
 to charge, $\rho({{\bf x}},t)$, and current, $\rho({{\bf x}},t){\vec v }$, densities of the extended electron. Contrary to Lorentz, the electron's charge distribution  was not postulated by Mie within a microscopic spatial volume (electron's fields were assigned by Lorentz to charge-free regions or to the supposed empty space around charges). The mathematical equation $div \ {\vec {d}}({{\bf x}},t) = 4\pi\rho({{\bf x}},t)$ can be rigorously resolved under the nonempty space assumption by admitting that the elementary charge and its field coexist together in all space points ${\bf x}$ of the infinite Universe. In other words, we  maintain that the Mie (and Einstein) idea that the elementary electric (and gravitational) charge is to be inseparably integrated into its spatial field structure with instantaneous local relations (yin-yang pairing).

The distributed radial electron with the analytical material density ${n}(r)>0$ possesses the same inertial mass density $m_o{n}({{\bf x}},t)> 0$ as the gravitational field mass density due to the Einstein's Principle of Equivalence.  This gravimechanical principle is a part of the Yin-Yang Principle for all natural phenomena. By following the Chinese dialectic, 
  the \lq yin\rq{} electric charge density $\rho_{yin}({{\bf x}},t) = q_{yin} n({{\bf x}},t)$ of the extended electron should be equalized with the paired charge density of local Coulomb fields, $\rho_{yang}({{\bf x}},t) = q_{yang} n({{\bf x}},t), q_{yang} = q_{yin}$. In the Einstein-type terminology,  the passive charge density $\rho_{p}({{\bf x}},t) = \nabla {\vec d} ({{\bf x}},t)/4\pi $ of the extended particle  
 numerically coincides with the active energy-charge density of the Coulomb field:
\begin {equation}
 \rho_a({{\bf x}},t)\equiv  
 [\pm \vec{d}({{\bf x}},t)]^2/4\pi\varphi_o,
\end {equation}
 where $\rho_{a}({{\bf x}},t) =\rho_{p}({{\bf x}},t)$ and  $\varphi_o=const$ is the universal self-potential for densities of active charges. Such physical relations between the particle\rq{}s density and its Coulomb\rq{}s  field intensity can be found \cite {Bul} as an exact solution to the Maxwell-Lorentz equations (1),
   \begin {eqnarray}
 \cases {
\rho (r)  = qr_q /4\pi  r^2(r + r_q)^2 =  \nabla {\vec{d}}(r) / 4\pi =  r_q{\vec{d}}^2(r) / 4\pi q \cr\cr
{\vec{d}}(r) = {q{\hat {\bf r}}/  r(r+r_q)} 
\cr\cr
{\cal E}_{q} = \int dV {\vec d}^2/4\pi  = q^2/r_q  \equiv \int\!dV \rho \varphi_o = q\varphi_o \ne \infty, \cr  
}\end  {eqnarray}
   in full confirmation that the Yin-Yang Principle equally works for paired active/passive densities of charges in Maxwell's electrodynamics and in Einstein's gravitation. 
      The active charge $q$ (of Coulomb fields with their energy density $ \rho\varphi_o$ and the strong field scale $|r_q| = |q/\varphi_o|$) has the energy nature under the universal potential $\varphi_o = const$ like an inertial  mass has the energy nature due to the celebrated formula $E= mc^2$.  Maxwell-Lorentz's equations  can  be discussed as in dual terms of electric fields and their currents, as well as in nondual terms of paired fields with continuous active/passive charge densities in the conjugated structure of active/passive energy streams.  In general, the self-energy $E_q= q\varphi_o$ of the active charge $q$ for electricity should be added to the  mechanical self-energy $E_m = mc^2$ of the active, inertial mass. Potential self-energies of passive charge densities  in their own electric/gravitational fields compensate quantitatively energies of active charges. These passive self-energies are infinite for point particles in Newton/Coulomb theories, but can be counted in the Cartesian world of extended elementary carriers of energy.

Now we associate electricity with imaginary numbers by using $\int\!\! \rho ({{\bf x}},t)dv \equiv q \equiv ie \Rightarrow - ie_o$ 
for the \lq{}negative\rq{} elementary charge of the electron. Its self-potential $\varphi_o \equiv  E_q/ q = q/ r_q = c^2/{\sqrt G}$ is to be 
 defined via the light speed limit $c$ and the Cavendish constant $G$. The similar charge to spatial scale ratio, ${\sqrt G}m/r_m = c^2/{\sqrt G}$, arises in  gravitation of continuous radial masses \cite {Bul, BulR}, where  $mc^2 \equiv {\sqrt G}m \varphi_o $ is the body internal (kinetic or active) energy. 
In other words, we use the fundamental potential $\varphi_o \equiv c^2/{\sqrt G} = 1.04 \times 10^{27} V$ for self-energies of both mechanical (real) and electric (imaginary) charges,
\begin {equation}
E = ({\sqrt G}m + q)\varphi_o =    i(e - i{\sqrt G}m) \frac {c^2}{\sqrt G},
\end {equation}
in order to extend the Einstein mechanical formula to electricity \cite {Buly}.  

Despite  that the electron possesses  in (4) the imaginary electric energy ($ - i\cdot 1,04 \times 10^{24} KeV$ of the imaginary charge $q =-ie_o$) next to 511 $KeV$ of the real mechanical energy, paired interactions of imaginary electric charges correspond to real Coulomb forces and to real  interaction energies. Contrary to an electric charge defined in real numbers, like the inertial mass, the imaginary electric charge exhibits a correct direction of the radiation self-force $2q^2{\ddot {\vec v}}/3c^3 $, which is proportional to $q^2 = (-ie_o)^2 < 0$. Ultimately, joint densities of mechanic and electric charges can be described in complex functions. Real integrals over spatial densities provide inertial mass-energies, while imaginary integrals correspond to non-inertial self-energies of electric charges.

\bigskip
\noindent {\bf Superimposition of extended charges with conservation and compensation of active/passive field energies} \smallskip

The  strong-field potential $W({\bf x}) = - \varphi_o ln [1 +{\sqrt G}m/ |{\bf x} | \varphi_o]  \neq const $ within the extended gravitational/inertial charge ${\sqrt G}m$ was introduced \cite {Bul} for local action on probe bodies  and for the local gravitational self-action of  massive densities themselves. In this view, the inertial (active, positive, kinetic) energy $\varphi_o {\sqrt G} m = mc^2 >0$ of the active charge ${\sqrt G} m$ is accompanied by  the negative gravitational (passive, negative, potential) self-energy $(-mc^2)$ because of the gravitational action of the  negative potential $W({\bf x})$ on its positive mass-charge densities ${\sqrt G}\rho_m ({\bf x}) = {\sqrt G}m r_m/4\pi {\bf x}^2 (r_m + |{\bf x}|)^2 $. Computations provide exact compensation of active (inertial, measurable) and passive (gravitational, unmeasurable) energy integrals for every radial (elementary) carrier of inertia and gravitation,
\begin {equation}
\varphi_o {\sqrt G} m \equiv    \int \varphi_o {\sqrt G}\rho_m ({\bf x}) d^3x = -   \int  W ({\bf x}) {\sqrt G}\rho_m ({\bf x}) d^3x = mc^2. 
\end {equation}
 This universal compensation of  the internal (kinetic, active) rest-mass energy $mc^2$ by the negative potential energy from the gravitational self-action,  $\int [\varphi_o + W ({\bf x})]{\sqrt G}\rho_m d^3x \equiv 0$, takes place under the integral mathematical equality $\int_o^\infty dx [ln (1 + x^{-1})]/(1+x)^2$ $\equiv \int_o^\infty dx/(1+x)^2$ $ = 1$.
Similarly, radial electric densities in their strong-field logarithmic potential $W ({\bf x})$ should  also lead to compensation of their post-Coulomb field energy. The latter and the potential self-energy are both infinite  in the point charge model, but they are finite for strong fields of the extended radial charge in consideration.

The gravity-type Poisson equation,  $\nabla^2 W = 4\pi {\rho}   \Rightarrow \varphi_o^{-1} (\nabla W)^2  $, treats the imaginary (astro)electron as a non-linear field composition with respect to the radial field intensity ${d}(r)$ or the electron\rq{}s interaction potential $W(r)$,
 with ${d}(r) = -\partial_r W (r)$ and 
\begin {eqnarray}
\frac { \partial_r [r^2 \partial_r W(r)] }{4\pi r^2}\equiv  \frac { [\partial_r W (r)]^2} {4\pi \varphi_o}=  {\frac {q r_q} {  4\pi r^2 (r + r_q)^2}}.  
\end {eqnarray}
 This non-linear equation reveals the imaginary post-Coulomb potential with the negative,  gravitational sign for the continuous radial carrier of electricity,
\begin {equation}
W (r) = - {  \frac {q}{ r_q}} ln  \left (1 + {\frac {r_q}{ r}} \right ) \equiv - {\varphi_o } ln  \left (1 + \frac {q}{ \varphi_o r} \right ).
  \end {equation}

The logarithmic potential (7) reproduces the Coulomb law on measu\-rable scales,  $(- q/r_q) ln [(r + r_q)/r] $ $\approx$ $ (-q/r) $ for $|r_q| \ll r$, but with the Newtonian negative sign. Such a  Newton-type potential of imaginary charges describes 
real electric forces, 
  $q_1 [-\nabla (-q_2r^{-1}_q)ln (1+ r_qr^{-1})] =  ie_1(-ie_2){\hat {\bf r}}/ r(r+r_q)$ 
$\approx e_1e_2{\hat {\bf r}}/ r^2$, with the mutual repulsion of like charges $e_1$ and $e_2$, and the attraction of unlike ones.

The gravitational sign in the Poisson equation for imaginary electric energies suggests to inverse positive and negative imaginary densities of  electric charges, $\rho \rightarrow -\rho$, in the Maxwell-Lorentz equations (1). The conventional inversion of electric charge signs for electrons and protons does not change electrodynamic laws of paired interactions. 
The electrostatic solution for charged imaginary densities in (1) with  the gravitational type direction of the radial Coulomb field, ${\vec {D}}({\bf x}) =  - {Q {\hat {\bf x}} } / {|{\bf x}|(|{\bf x}|+Q \varphi_o^{-1})}$,  facilitates the unification of gravity and electricity on the basis of one  complex energy - charge (4), $Q\varphi_o \equiv (q_m+iq_e)c^2/{\sqrt G} \Rightarrow ({\sqrt G} m + ie)c^2/{\sqrt G}$. The Maxwell-Lorentz equations (1) for the massless  elementary charge can be extended to the complex energy flows of the radial carrier  with the elementary charge density $\rho ({{\bf r}},t) \equiv ({\sqrt G} m + ie)n ({{\bf x}},t)= - div\ {\vec {D}}({{\bf x}},t)/ 4\pi$,
\begin {eqnarray}
 \cases {
  div\ {\vec B}({{\bf x}},t)\varphi_o = 0 \cr
 c\ curl\  {\vec B}({{\bf x}},t)\varphi_o  = {\vec v}div\ {\vec {D}}({{\bf x}},t)\varphi_o + \partial_t  {\vec {D}}({{\bf x}},t)\varphi_o\cr
c\ curl\ {\vec {D}} ({{\bf x}},t) \varphi_o = - \partial_t {\vec B}({{\bf x}},t)\varphi_o.\cr
} 
\end {eqnarray} 
Here we replaced the extended charge density $\rho ({{\bf r}},t)$ in the nondual Maxwell-Lorentz equations with the equivalent density 
of the local field divergence. The yin-yang field physics of locally paired material densities maintains nondual energy reality in classical  electrodynamics in a line of electrodynamics  for nondual quantum fields.

By locally pairing energy flows of the radial Newton-Coulomb fields and the densities of the continuous particle, one also unifies in (8) gravi-mechanical and electric self-energies $q\varphi_o \Rightarrow Q\varphi_o =  ({\sqrt G} m + ie) c^2/ {\sqrt G} $. Such a double unification (particle with field and gravity with electricity) requires one (but complex) charge $q\Rightarrow Q = ({\sqrt G} m + ie)$ in the  logarithmic
   potential (7).   One can verify quantitatively that the complex (astro)charge distribution, $\rho(r)=Qn(r) = ({\sqrt G} m + ie)n(r)$ with $r_q \Rightarrow z_Q = Q{\sqrt G}/c^2$,  generates  this strong field potential under the regular integral  rule for classical gravitation and electrodynamics,
\begin{eqnarray} 
\!\int\!\frac{(-Q) n({r}')dv'}{|{\vec r}-{\vec r}'|}
=\!-\!\int_o^\infty\!\int_{-\pi/2}^{\pi/2}\!\int_o^{2\pi}\!\!\frac {d\phi'sin\theta'd\theta'r'^2dr'}{\sqrt {r^2\!+\!r'^2\!-\!2rr'cos\theta'}}\frac{Q z_Q}{4\pi r'^2(r'\!\!+\!\!z_Q)^2} \cr
= -  \int_o^\infty\! \frac {dr' Qz_Q}{  (r' + z_Q)^2} \left(\frac  {{ {|r' + r|} }\!-\!{ {|r' - r|} }}{ 2rr'} \right )  
 = - 
 \!\!  \int_r^\infty\!\!  \frac{Qdr'}{ z_Q} \left (\frac{1}{r'}\! -\! \frac{1}{r'+z_Q}  \right)  \cr
 \! \equiv\! \frac {(-Q)}{z_Q} ln \left(1+ \frac {z_Q}{r} \right)\!\! =\!\!\int_r^\infty\!\! {{\hat {\bf r}}\rq{}{\vec D}(r')dr'}=W_{Q}(r) .
\end{eqnarray} 
 Notice from the last relation in (9) that $ W_{Q}({r})$ coincides with the work associated with  replacements of a unit probe charge from the point 
 ${r}$ to  $\infty$ against the radial field  ${\hat {\bf r}}\rq{}{\vec D}(r') = D(r) = - \partial_r W(r)$. The integration over $r'$ within $0 \leq r' \leq r$  vanishes identically in (9) in agreement with the physical meaning of potentials for a probe body.

 The  multibody system of complex charges $Q_k= {\sqrt G}m_k + ie_k\equiv \varphi_o z_k \equiv {\cal E}_{k}/\varphi_o$ with strong field electric and gravitational interactions can be described in statics by the \lq long\rq{} logarithmic potential: 
 
 \begin{equation}
W_{sys}({\bf x})	= - \frac {c^2}{\sqrt G} ln \left (\!1\!\!+\!\!\frac {z_1}{|{\bf x}\!\!-\!\!{\bf a}_1 |}\!+\!\frac {z_2}{|{\bf x}\!-\!{\bf a}_2 |}\!+...+\!\frac {z_n}{|{\bf x}\!-\!{\bf a}_n |}\!\right ).
\end{equation}
Like in the equality (6), the Yin-Yang Principle universally works for net  active and passive densities of overlapping complex charges (or their self-energies) in all field points of  the common material space-plenum,
\begin{equation}{ \rho_a({\bf x})} \equiv \frac {[-\nabla W_{sys}({\bf x})]^2}{4\pi } = \frac {\nabla^2 W_{sys}({\bf x})}{4\pi{\varphi}_o} \equiv  {\rho_p({\bf x})}.
\end{equation}

In spite of paired interactions of elementary extended charges, their volume energy integral  demonstrates the system energy conservation law due to interference (dark, invisible) exchanges
$2AB$ in the algebra structure $(A+B)^2$ of overlapping energy densities,  
\begin{eqnarray}
{\cal E}_{sys}\equiv\! \int\! \rho_a({\bf x}){\varphi}_o d^3x\cr =  \frac {{\varphi}^2_o}{4\pi }\! \int 
\left (
 \frac {    \frac {{({\bf x} - {\bf a}_1)z_1}}{|{\bf x} - {\bf a}_1 |^3}    + \frac {({\bf x} - {\bf a}_2)z_2}{|{\bf x} - {\bf a}_2 |^3} +...+ \frac {({\bf x} - {\bf a}_n)z_n}{|{\bf x} - {\bf a}_n |^3}}{ 1 + \frac {z_1}{|{\bf x} - {\bf a}_1 |}     + \frac {z_2}{|{\bf x} - {\bf a}_2 |} +...+ \frac {z_n}{|{\bf x} - {\bf a}_n |} } \right )^2\!d^3x 
 \equiv 
 {\varphi}^2_o \sum_{k=1}^{n} z_k\equiv \sum_{k=1}^{n} {\cal E}_{k},
	\end{eqnarray}
	Again, mono-radial densities $A^2({\bf x} - {\bf a}_A)$ and $B^2({\bf x} - {\bf a}_B)$ can be interpreted under observations as elementary carriers of energy due to measurable radial structures of their  Coulomb/Newton fields. It is possible to extract these mono-radial carriers from a visible substance like its parts (radial particles).   Based on the experimental data, it can be assumed that only mono-radial elementary carriers can constitute  the visible substance. Indeed, bi-radial material densities $A({\bf x} - {\bf a}_A)B({\bf x} - {\bf a}_B)$ in the energy integral (12) cannot be in principle retrieved from the visible body together with one particle or one corpuscle. Thus, the interference (invisible, dark) energy densities can be justified  for  electrodynamics and gravitation only from calculations rather than from experiments and observations.	This justification contradicts the motto \lq Nullius in verba\rq{.}

	Another yin-yang universal law of this Cartesian physics of material fields is the exact compensation of Coulomb/Newton active energies (12) by the self-action of passive charged densities from (11) in their relativistic potential (10):    
		\begin{eqnarray}
 \int \!
\left (
 \frac {    \frac {{({\bf x} - {\bf a}_1)z_1}}{|{\bf x} - {\bf a}_1 |^3}    + \frac {({\bf x} - {\bf a}_2)z_2}{|{\bf x} - {\bf a}_2 |^3} +...+ \frac {({\bf x} - {\bf a}_n)z_n}{|{\bf x} - {\bf a}_n |^3}}{ 1 + \frac {z_1}{|{\bf x} - {\bf a}_1 |}     + \frac {z_2}{|{\bf x} - {\bf a}_2 |} +...+ \frac {z_n}{|{\bf x} - {\bf a}_n |} } \right )^2\!
  \frac {{\varphi}^2_o d^3x}{4\pi }\cr \!\!
-\! \int \!
\left (
 \frac {    \frac {{({\bf x} - {\bf a}_1)z_1}}{|{\bf x} - {\bf a}_1 |^3}    + \frac {({\bf x} - {\bf a}_2)z_2}{|{\bf x} - {\bf a}_2 |^3} +...+ \frac {({\bf x} - {\bf a}_n)z_n}{|{\bf x} - {\bf a}_n |^3}}{ 1 + \frac {z_1}{|{\bf x} - {\bf a}_1 |}     + \frac {z_2}{|{\bf x} - {\bf a}_2 |} +...+ \frac {z_n}{|{\bf x} - {\bf a}_n |} } \right )^2\!
\! ln \!\left (\!1\!\!+\!\!\sum_{k=1}^n\frac {z_k}{|{\bf x}\!-\!{\bf a}_k |}\!\right ) \frac {{\varphi}^2_o d^3x}{4\pi }\!
 \equiv 0 .
	\end{eqnarray}
	The yin-yang Universe with the mass/charge creation mechanism from \lq the void nothing \rq {} (i.e. from the net zero energy)  stands behind the global integral compensation (13) for active and passive densities of complex energies.  The Cartesian world of extended matter is a self-consistent one in energy terms contrary to the Newton empty space with material singularities, Coulomb energy divergence and astronomical black holes. 
	
\bigskip
\noindent {\bf Maxwell equations are quantitative identities for paired energy flows }
  
The exact analytical  solution (3) to Maxwell's equations rigorously requests the $r^{-4}$ radial astroparticle instead of the conventional point particle (or  the smallest material corpuscle assumed  by observers in microscopic volumes). However, rigorous solutions demand that 
the empty space paradigm (presumably resulting from observations and measurements) should be replaced with the non-empty space concept to achieve  a coherent theory of classical fields. Distributed elementary energies with smooth  densities should stand behind the yin-yang reality, which is available to everyone in observations of regions with high (former particles, substance) and low (weak fields) concentrations of energy.  

Like the first pair of Maxwell identities, $\nabla_\lambda F_{\mu\nu}+
\nabla_\mu F_{\nu\lambda} + \nabla_\nu F_{\lambda\mu} \equiv 0$, the second pair of Maxwell  equations for currents can also be considered (in the non-empty space paradigm) as quantitative identities for balanced momentums  ($\rho \varphi_o u^\mu /c$) of yin-yang paired energy densities, 
\begin {eqnarray}
 \cases {
[\nabla_\lambda F_{\mu\nu}(x)+
\nabla_\mu F_{\nu\lambda}(x) + \nabla_\nu F_{\lambda\mu}(x)]\varphi_o/4\pi c \equiv 0 \cr
[ \delta^\mu_\lambda - u^\mu(x) u_\lambda (x)] \nabla_\nu F^{\nu\lambda}(x)\varphi_o/4\pi c\equiv 0. \cr }
\end {eqnarray}
Here we used the following identical relations for classical fields:
\begin {eqnarray}
 \cases {
\nabla_\nu F^{\nu\mu}(x)\equiv  4\pi j^\mu(x)/c \equiv 4\pi \rho(x) u^\mu(x) \cr
u_\mu(x)\nabla_\nu F^{\nu\mu}(x) \equiv 4\pi \rho(x) \cr
\nabla_\nu F^{\nu\mu}(x) \equiv   u^\mu(x) u_\lambda (x)\nabla_\nu F^{\nu\lambda}(x). \cr
} 
\end  {eqnarray}

  Energy-momentum balance of the pair of 4-vector flows in the second identity in (14) requires the absence of  energy currents along directions normal to the 4-velocity $u^\mu$ of the field-energy density (with $u_{\mu}u^\mu = 1$). There is no sense in the extra particle notion under the yin-yang  scheme of locally paired field densities in Maxwell identities. In other words, the material  reality is nothing but inhomogeneous energy flows in all spatial points of the Universe, filled continuously with densities of visible  (corporeal)  and invisible  (incorporeal) matter.

If  we consider the electric currents through the coupled field flows of energy in the Maxwell-Lorentz theory for the extended electron, then  the variation techniques for paired material fields in unified physics of nonempty space must be different from the textbook variation procedures for the empty space physics of spatially separated particles and fields. Indeed, the current Maxwell equations were derived in the dual action approach to fields and remote particles (which can be varied and fixed independently, like $\delta [A_\mu j^\mu] = j^\mu \delta A_\mu$). In the yin-yang unified approach, self-energy variations of  locally paired flows are coupled inseparably and  $\delta (A_\mu j^\mu)$ is equal to the sum $j^\mu \delta A_\mu + A_\mu \delta j^\mu$ that doubles the variational result, 
\begin {eqnarray} \delta (cA_\mu\nabla_\nu F^{\nu\mu}/4\pi) =(c/4\pi) [\delta A_\mu \nabla_\nu F^{\nu\mu}   + A^\mu\nabla^\nu (\partial_\mu \delta A_\nu - \partial_\nu\delta A_\mu ) ]\cr
\Rightarrow  (c/2\pi) \nabla_\nu F^{\nu\mu}  \delta A_\mu = 2j^\mu  \delta A_\mu. \end {eqnarray}
Thus, the yin-yang Lagrangian for locally paired energies should possess the double electromagnetic density,  $2\times F_{\mu\nu}F^{\mu\nu}/16\pi$, of charged material fields, \begin {equation}
S_{y-y} = - \frac {1}{c} \int d \Omega \left [ \frac {1}{c}A_\mu j^\mu  +  
 \frac {1}{8\pi} F_{\mu\nu} F^{\mu\nu} \right],
\end {equation}
compared to the conventional alternative with spatially separated fields and charges.
The  yin-yang paired variations of (17) with respect to both potentials $A_\mu$ and currents $\j^\mu \propto A^\mu$ result in the  Maxwell relation  $4\pi j^\mu / c = \nabla_\nu F^{\nu\mu}$ or the energy flow identities in  (14). It is worth noting that the yin-yang action (17) vanishes for  Lagrange variation trajectories:
 \begin {eqnarray}
S_{y-y} = - \frac {1}{c} \int d \Omega \left [ \frac {1}{c}A_\mu j^\mu  +  
 \frac {1}{8\pi} F_{\mu\nu} F^{\mu\nu} \right] \cr = -  \int \frac {d \Omega}{8\pi c} \left [ 2 {A_\mu \nabla_\nu F^{\nu\mu} } +  
 ( \nabla_\mu  A_\nu -  \nabla_\nu A_\mu ) F^{\mu\nu} \right]
\cr 
 =  -  \int \frac {d \Omega}{4\pi c}  A_\mu \nabla_\nu (F^{\nu\mu} + F^{\mu\nu}) +
 \oint \frac {dS_\nu}{4\pi c}  A_\mu F^{\mu\nu} =-0+0.
 \end {eqnarray}
The integral nullification of charge-energy currents $\varphi_o j^\mu$ in (18) is in full agreement with the yin-yang compensation law for the case of static matter (13), where the active energy of extended charges is balanced by the passive  potential energy due to the self-action in their post Coulomb/Newton radial fields.
 In other words, the action nullification on the Lagrange trajectory is the  expected result for the yin-yang nondual physics.  We recall that the conventional current + field action in dual physics, where  ${\bf d}^2/8\pi$ is instead of ${\bf d}^2/4\pi$ in (2), and $F_{\mu\nu}F^{\mu\nu}/16 \pi$ instead of $F_{\mu\nu}F^{\mu\nu}/8 \pi$ in (17), does not match  such a Lagrange path nullification.  
  This obvious mathematical flaw is the penalty for using an empty space model and for its misinterpretation of continuous material reality. Illogical changes in magnetic fields at fixed currents that locally and identically generate these fields, including any changes of them, have always been one of the unsolved problems of Classical Electrodynamics.

\bigskip \noindent {\bf 1938 Einstein's material fields for 1629 Cartesian vortex mechanics} \smallskip

The evolution of Einstein's theory of relativity has already passed three milestones. They are the first two postulates and the active internal energy $mc^2$ of the 1905 Special Relativity, the geometrization of massless metric fields in the 1915 Einstein Equation under the Newton's empty space dogma, and the 1938 proposition to distribute particle's mass-energy $mc^2$ continuously over all spatial points of the material metric field in the non-Newton nondual approach to physical reality.  Recall that the integration of particles into spatial structures of their fields was suggested by Einstein together with Infeld  \cite {Ein} for the further evolution of all physics disciplines: \lq\lq We would regard matter as being made up of regions of space in which the field is extremely intense... There would be no room in this new physics for both field and matter, for the field would be the only reality.\rq\rq{} However, the extended mass  has not been yet adopted by modern relativists.  Their Newtonian references traditionally associate  General Relativity (GR) only with the 1916 empty space metric of Schwarzschild. And they use  the Dirac delta-function for the formal presentation of  the  Lagrange material density $\mu({\bf x}) = m \delta ({\bf x}-{\mbox{\boldmath$\xi$}})
       / {\sqrt {\gamma}}$ 
       in the point mass action  $S = - c\int\!\!\int\!\!\int\!\!\int\!\!{\sqrt {\gamma}}\mu({ \bf x}) ds d^3x = - c\int m ds({\mbox{\boldmath$\xi$}})$ for the Newton empty space $x^i$ with material peculiarities along the path ${\mbox{\boldmath$\xi$}}(x^o)$.  But, the mass density is to be a continuous space-time function in the Cartesian world and the Einstein-Infeld pure field approach to matter \cite {Ein}. The best mathematical candidate in pseudo-Riemannian geometry to match this scalar physics function $\mu(x)$ is the Ricci scalar density $R(x)$. Indeed, the latter is also unique for a particular distribution of mass-energy in  pseudo-Riemannian space-time. By following the Einstein and Infeld propositions of 1938 for nonempty space reality of massive fields,  one  may request  linear proportionality of  physics and mathematics scalar densities, $\mu(x) \equiv  \zeta \varphi_o^2 R(x)/ c^2 $. Here again $\varphi_o = c^2/ {\sqrt G} = 1.04\times 10^{27} V$ is the universal potential for the inertial/gravitational energy charge $q_m\equiv E_m/ \varphi_o ={\sqrt G}m $ and $\zeta$ is the dimensionless coupling constant (subject to be defined).  

Cartesian vortex dynamics requires to distribute 
 the \lq rest\rq{} mass-energy $mc^2$ continuously over all GR 4-volumes  ${\sqrt {-g}}d^4x \equiv {\sqrt {g_{oo}}dx^o } {\sqrt \gamma} d^3x$ in the action integral: 
 \begin {equation}
S= -c \int\!\!\int\!\!\int\!\!\int\!\!  \mu (x) ds(x){\sqrt \gamma} d^3x = 
 -  \frac {\zeta \varphi^2_o}{c} \int\!\!\int\!\!\int\!\!\int\!\!  R(x)\frac  {ds(x)}{{\sqrt {g_{oo}}dx^o }} {\sqrt {-g}}d^4x.
\end {equation} 
Therefore, the Cartesian approach to nature reveals the kinematic scalar field $B=ds/{\sqrt {g_{oo}}}dx^o $ next to the inhomogeneous Ricci scalar $R$  in  the nonempty space-time with inertial densities,
\begin {equation}
B(x) \equiv \frac {\sqrt {g_{\mu\nu}dx^\mu dx^\nu }} {\sqrt {g_{oo}}dx^o }
\equiv \frac {1}{{\sqrt {g_{oo}}}u^o}
\equiv \frac {\sqrt {1-\beta^2}} {1-  g_{oi}v^i / {\sqrt {g_{oo}}} }
\equiv  {\sqrt {1-\beta^2}} \left (1 + \frac { g_{oi}dx^i }{  {g_{oo}}dx^o    }
\right ).
\end {equation} 
 This inertial field depends on the GR metric tensor $g_{\mu\nu} = g_{\nu\mu}$, the physical three-velocity $v^i \equiv c dx^i/ {\sqrt {g_{oo}} dx^o} [1+ (g_{oi}dx^i / g_{oo}dx^o)]  $ and the relativistic speed factor $\beta^2 \equiv \gamma_{ij} v^iv^j/ c^2$, with $\gamma_{\mu\nu} \equiv  g_{o\mu}g_{o\nu}g^{-1}_{oo}  - g_{\mu\nu}$. The Hilbert-type variations of (19) with respect to $\delta g^{\mu\nu}$, namely $\delta S =\int \delta g^{\mu\nu} T_{\mu\nu} {\sqrt {-g}} d^4x/2c$, define the energy tensor density of moving inertial/gravitational fields,  
 \begin {equation}
T_{\mu\nu} \equiv  {\zeta \varphi_o^2} \left [(u_\mu u_\nu - \gamma_{\mu\nu}){BR} - 2BR_{\mu\nu} + 2{\nabla_\mu \nabla_\nu B - 2g_{\mu\nu} \nabla_\lambda \nabla^\lambda B }
\right ],
\end {equation}
associated to the covariant material 4-flows, $\zeta \varphi_o^2 R u_\mu  \equiv \varphi_o^2 \zeta R g_{\mu\nu}dx^\nu/ds$,  of continuous masses.
 
The kinematic field  (20) is constant, $B=1$, for the static case, when $v^i=0, \beta^2 =0$, 
 $u_\mu = \{ {\sqrt {g_{oo}}}; \   g_{oi}/{\sqrt {g_{oo}}} \}$, $u_\mu u_\nu - \gamma_{\mu\nu} = g_{\mu\nu}$, and $T_{\mu\nu}\Rightarrow  {\zeta \varphi_o^2} (g_{\mu\nu} R - 2R_{\mu\nu}) \equiv T^{static}_{\mu\nu}$. The latter is the  mass-energy tensor of static material densities in  the 1938 pure field theory  of Einstein and Infeld. Again, in 1915 Hilbert and Einstein considered field densities as massless and, therefore, motionless. They separated the static gravitational field from moving inertial particles  by replacing the Ricci tensor construction $(g_{\mu\nu}R - 2R_{\mu\nu})c^4/16\pi G$ with the opposite sign to the left-hand side of the 1915 Einstein Equation. Indeed,   Hilbert varied the massless Ricci density $R$ without any options for its motion in the 1915 field action by formally taking $Rds \rightarrow R{\sqrt {g_{oo}} dx^o}.$  Mathematics  is able to study only the coordinate field contribution to the  Lagrange dynamical equations in the empty space  model of Newton. By following Einstein and Infeld, we admitted the medium-like motion of field matter in (19). And we related the geometrical curvature $R$ to inertial mass-energy flows $Ru_\mu$, which are, in fact, metric space-time flows of the material space-plenum.

Cartesian natural philosophy considers the non-stop vortex motion as an  origin of matter and its energy. Therefore, the steady vortex auto-organization could be considered as an equilibrium state of the elementary extended mass in the Einstein-Infeld theory of nondual material reality.      
   The geometrical 4-flow of matter  provides vanishing spatial transport of mass-energy under the metric equilibrium,
   \begin {equation}
R_{eq}u^{eq}_\mu \equiv
\left \{ 
 R_{eq}\frac {\sqrt {g^{eq}_{oo}}     } {\sqrt {1-\beta^2_{eq}}} ; \ \ 
R_{eq}\frac { (g^{eq}_{oi}/{\sqrt {g^{eq}_{oo}}} ) - v_i^{eq}/c }{\sqrt {1-\beta^2_{eq}}}
\right \} = \{R_{eq}; \ \ 0  \}, 
 \end {equation}    
      where $v^{eq}_i / c =   g^{eq}_{oi}/{\sqrt {g^{eq}_{oo}}} = \gamma_{ij} g_{eq}^{oj}{\sqrt {g^{eq}_{oo}}}= \gamma_{ij}v_{eq}^j $,       $v_{eq}^j /c = {\sqrt {1-\beta^2_{eq}}} (g^{\mu j}u_\mu) = {\sqrt {g^{eq}_{oo}}}  g_{eq}^{oj} $,      
      ${\sqrt {g^{eq}_{oo}}     } /{\sqrt {1-\beta^2_{eq}}} = u^{eq}_o = u^o_{eq} =1$, $g^{eq}_{oi}g_{eq}^{oi} =\beta^2_{eq} = 1- g^{eq}_{oo}$, and $B_{eq}= 1/{\sqrt {g^{eq}_{oo}}} = 1/{\sqrt {1-\beta^2_{eq}}} $.
       There are no stresses within  the elementary  material space in its dynamical equilibrium  and the energy-tensor   density (21)  is to be traceless in all points of the elementary Cartesian vortex (22), 
 \begin {equation}
g^{\mu\nu}_{eq} T^{eq}_{\mu\nu} \equiv 
 {2\zeta \varphi_o^2 } \left ( B_{eq}R_{eq} -  3\nabla_\lambda \nabla^\lambda B_{eq} 
        \right ) = 0.
 \end {equation} 
 
            The radial auto-organization of the  elementary mass-energy $mc^2$ within its continuous material space is accompanied by six inherent geometry symmetries \cite{Bul} ($\gamma_{ij} = \delta_{ij}$, with $g^{ij} = - \delta^{ij}$, ${\sqrt {-g}/{\sqrt {g_{oo}}}} \equiv {\sqrt \gamma} =1,$  $g^{oi} =  \delta^{ij}g_{oj}/g_{oo} $)  under the radial distributions of the continuous mass-energy, $\mu(r)c^2 = m c^2r_m/4\pi r^2 (r+r_m)^2$,  and the time related  metric component $g_{oo} = r^2/(r+r_m)^2$, $r_m \equiv Gm/c^2$. This steady radial distribution of elementary matter is also a stationary solution of the Klein-Gordon type equation (23), 
 \begin {eqnarray}
 {\zeta \varphi_o^2 R_{eq}}
 = \frac {3\zeta \varphi_o^2{\sqrt{g^{eq}_{oo}}}  } {{\sqrt {-g_{eq}}}  } {      \partial_\mu \left [{\sqrt {-g_{eq}}} g_{eq}^{\mu\nu}\partial_\nu \left ( \frac {1}{\sqrt {g^{eq}_{oo}}} \right )\right ] } \nonumber \\ = -  {3\zeta \varphi_o^2 }\delta^{ij}\partial_i\partial_j ln \frac {1}{{\sqrt {g^{eq}_{oo}(r)}}}
\Rightarrow m c^2\frac {r_m}{4\pi r^2 (r+r_m)^2},
 \end {eqnarray}         
      that here defines the coupling constant $\zeta$ as $1/12\pi$. 
      
      The stationary radial solutions of (22) - (24),  $ u_o(r) = 1, u^o(r) = 1, u_i(r) = 0,  u^i(r) \neq 0,  v^2(r) = r_m/(r+r_m),$  and $B_{eq}(r)= 1/{\sqrt {g^{eq}_{oo}(r)}}= (r + r_m)/r$ for the equilibrium vortex motion, mean that the geometrical Ricci scalar in (24) originates from the inertial field $B_{eq}\neq const$. 
     In other words, the elementary vortex  mass $m = r_m c^2/G$ has the pure kinematic, Cartesian origin due to $v^2(r) \neq 0$ for the equilibrium creation of elementary inertial and gravitational energies. The fact that the initial cause of everything lies in mechanical motion was understood by Descartes in the 17th century. Chaotic vortex motions of Ricci metric densities  are indeed responsible as for the kinematic creation of the internal mass-energy $mc^2$ within the elementary mater-extension of Descartes. It is essential that the vortex self-organization of the inertial (positive) mass-energy  is accompanied and balanced by the metric gravitation,   $g_{oo}(r) = 1- \beta^2(r) \neq const$, within the same material space. It is clear from the kinematic nature of the  Cartesian vortex that the internal mass-energy $mc^2$  of elementary matter   has the meaning of its internal relativistic heat, associated with chaotic vortex motions of metric densities at high orbital speeds. 
     
     Now we look again at the yin-yang compensation (13) of world energies from the Cartesian physics position. From where did the positive (kinetic) rest-mass energy emerge under the vortex auto-organization of the Cartesian matter-extension? The point is that, aside from the equilibrium  metric motion for the kinematic creation of inertial  mass-energy $\mu_{in}c^2$,  the  gravitational self-action always arises due to the intense interaction potential \cite{Bul} $W(r) \equiv \varphi_o ln {\sqrt {-g}} = -\varphi_o ln [(r+r_o)/r]  $ of the emerged charge density ${\sqrt G}\mu_{gr}$, with $\mu_{gr}\equiv \mu_{in}$:
  \begin {equation}
E_{kin}+U_{self} =  \int \left [\mu_{in}(r)c^2  - {\sqrt G}\mu_{gr}(r)\varphi_o ln \frac {1}{\sqrt {g_{oo}(r)}}\right ]d^3 x = (m_{in} - m_{gr}) c^2 \equiv 0. 
 \end {equation}      
 Contrary to measurable exchanges of  internal and translation kinetic energies, the negative  potential energy $ (- m_{gr}c^2)$ is not relevant to observations in practice.  But negative energy tensions within the continuous carrier of  the elementary relativistic heat is relevant to stability of the elementary extended mass-energy by the internal negative pressure, once assumed by Poincare for the extended particle.

      \bigskip \noindent {\bf Cartesian references for the Einstein Equation}
    \smallskip

     The internal heat, or an internal kinetic degree of freedom under the spatial transport of elementary energy,  cannot be reasonably assigned to the Newton point mass.  Therefore, the variable vortex heat is the principle difference between Newton and Cartesian transport of energy, even at the low speed motion of mechanical bodies. Indeed, the Newton point particle possesses only one degree for the summary (internal and translation) kinetic energy, while the Cartesian distributed vortex  
      at rest  for the elementary heat-energy possesses both kinetic internal and kinetic translation energies. The Special Relativity kinematic cooling $Q(\beta^2) = Q_o {\sqrt {1-\beta^2}} \approx (mc^2 - mv^2/2 - mv^4/8c^2 - ...)$ of the elementary rest-mass energy $ mc^2 \equiv Q_o$, which is now the elementary relativistic heat, and the pure translation (mechanical) energy $p_iv^i \equiv H(\beta^2)+L(\beta^2)
      =  Q_o/ {\sqrt {1-\beta^2}} - Q_o {\sqrt {1-\beta^2}} $ $ \approx (m v^2+mv^4/2c^2 +...)$ reitirate together the Newton-type summary changes of  internal (heat) and external (mechanical, translation) kinetic energies, $\Delta [Q(\beta^2) +p_iv^i] = \Delta (mv^2/2 + 3mv^4/8c^2 +...)$  for the low speed  ($\beta^2 \ll 1$) transport  of the elementary energy carrier with metric auto-rotations. 
      
      The slow spatial motion of the Cartesian  mechanical body  is first of all the transport of the variable heat integral $Q_o(1-\beta^2/2)$ of vortex densities. The  accompanying  fraction of the kinetic energy $Q_o\beta^2$ is counted as a small correction of chaos due to the translation ordering. Cartesian transport of two variable energies just formally correspond to the Newton transport of the constant mass $Q_o/2c^2$ plus one variable energy $Q_o\beta^2/2$.  Newton had no idea about the internal degree of kinematic heat-energy and modeled  transport energy changes,   $[Q_o(1-\beta^2/2) + Q_o\beta^2 ] - Q_o = Q_o
     \beta^2/2$, through one summary coefficient $Q_o/ 2 c^2 \equiv m/2$.  
     It is clear from Cartesian physics that Newton's model works properly only for small probe bodies in empty space where internal and external (translation) kinematic degrees of freedom obey the collinear transport laws. 
     
     In condensed media, where heat energy gradients and mechanical momentum may have different directions, Cartesian mechanics will not reiterate Newtonian  one. Moreover, the Newton model of motionless point masses without internal heat does not coincide with the  Cartesian system of static vortices which initially have internal kinetic energy or heat.  One can expect some principle differences between the dual and nondual theories of matter for its spatial acceleration and its heat transfer in liquids and gases.  Here, different exchange mechanisms can control non-Newtonian energy gains of $ mv^2 $ and thermal losses $ -mv ^ 2/2 $ due to kinematic cooling. 
     Early or later, the elementary relativistic energy of internal vortex motions with variable heat should  replace the Newtonian constant mass as a basic notion for description of  warm material space in General Relativity. The latter can not rely on Newton's cold masses in order to incorporate thermodynamics at the low speed limit. Equally, metric gravitation of cold masses cannot considered as a true limit for gravitation of Cartesian energy-charges, which depend on variable heat.

 The pure field physics of Einstein and Infeld claims mutual interactions of overlapping mass-energy densities of elementary heat structures, which are forming the indivisible whole called the warm Universe. The world material medium in any considered region is always out of  equilibrium, unless the global system has not reached some steady kind of self-organization for local material densities  in consideration. The tensor densities (21) for  nonequilibrium  material flows are balanced in the most general case by external stresses $P_{\mu\nu}$. This general balance for moving nonempty space in nondual physics of  Einstein and Infeld corresponds to an analog of the Einstein Equation for formally separated empty space and matter,
   \begin {equation}
 {\zeta \varphi_o^2 } \left [(u_\mu u^\nu - g^{\nu\lambda}\gamma_{\mu\lambda})\ {BR} - 2BR^\nu_{\mu} + 2{\nabla_\mu \nabla^\nu B - 2 \delta ^\nu_{\mu} \nabla_\lambda \nabla^\lambda B }
\right ] + g^{\nu\lambda}P_{\mu\lambda} = 0.
\end {equation}
    
    Einstein was the first researcher who tried to derive the motion of particles directly from his gravitational field equation. He was engaged into this conceptual problem for more than 20 years. Empty space regions in Newtonian dual physics are described by field solutions with zero Ricci curvatures, $R_{\mu\nu}=0$. The motion of point peculiarities in such massless fields can be approximated only through successive iterations \cite {AMH, AM, CAN}. Cartesian physics of material fields is free from these complicated iteration procedures. 
        One     
    can  find the exact  geodesic  equations for any non-equilibrium motion of  inertial Ricci densities in the world metric space-time  as the vector flow conservation from the   general tensor  balance (26), $\nabla_\nu (T^\nu_\mu + P^\nu_\mu)= 0$. Here it is sufficient to make use of the known geometrical equalities $\nabla_\nu \nabla_\mu \nabla^\nu B - \nabla_\mu \nabla_\nu \nabla^\nu B \equiv R_{\mu\nu} \nabla ^\nu B\equiv R_{\mu}^{\nu} \nabla _\nu B$ and   the Bianch identities 2$\nabla_\nu R^\nu_\mu \equiv \nabla_\mu R$:
         \begin {equation}
  \zeta \varphi^2_o  [\nabla_\nu ( BR u_\mu u^\nu ) - B\nabla_\mu R - \gamma_{\mu\lambda}\nabla^\lambda ( BR)] = - \nabla_\nu P^\nu_\mu .
  \end {equation} 
    This general equation of motion can be equally read for the 4-acceleration $u^\nu \nabla_\nu u^\mu$ along a normal axis to the 4-velocity $u^\mu$ of the scalar mass density $\mu = \zeta \varphi^2_o R/c^2$,  
          \begin {equation}
   B\mu c^2 u^\nu \nabla_\nu u^\mu  
   +  (u^\mu u^\nu - g^{\mu\nu} )[ \frac { g_{o\nu}}{g_{oo}}\nabla_o (\mu B)   -  \mu  \nabla_\nu B]c^2
      =   (u^\mu u^\nu - g^{\mu\nu} )\nabla_\lambda P^\lambda_\nu.
      \end {equation}
    The vector equation (28) for Cartesian mechanics claims that the inertial field $B(x) \neq const$ contributes to geodesic dynamics of material space densities, while dual theories with the empty space notion are free from such kinematic feedbacks.

    \bigskip \noindent {\bf Practical benefits from  Cartesian physics of Einstein and Infeld}
    \smallskip
    
     It is very difficult to infer from observations and measurements whether the world space is empty, as Newton modeled for localized bodies and point particles, or this space is filled continuously everywhere with invisible matter, as Aristotle and Descartes claimed.
 The experimental procedure of  a theory verification  can reject  improper approximations of reality, but cannot select the most correct approach among suitable alternatives \cite {Pop}. For instance, there are about ten competing formulations of Quantum Mechanics at the moment. General Relativity of 1916 was requested to reiterate Newton's empty space dogma. But 
Einstein himself began to share the Cartesian matter-extension for the pure field physics at least since 1938. And non-empty space metric constructions for Einstein's General Relativity lead again to the same post-Newton findings for the main gravitational tests \cite {BulR} as the empty-space metric of Schwarzschild can propose. What is the use of reinforcing  the  Aristotle-Descartes material space if it might not be ever justified in practice?

     Recall again that Newton's low speed dynamics works satisfactorily for small probe bodies where transports of the internal kinetic energy or relativistic heat $Q=mc^2(1-\beta^2/2) $ and the kinetic energy of spatial translations $mc^2\beta^2$ have collinear directions.  Newtonian proponents, who have to consider the mechanical  motion as a transport of the constant elementary mass $m = const$, derive dynamical laws for acceleration of such masses and, basing on these laws for constant masses, compute accompanying energy flows. 
          Cartesian physicists have to consider first of all the transport of elementary vortex energies with the kinematic cooling under motion, $mc^2(1-\beta^2/2) \neq const$,  and have to derive dynamical laws for vector energy flows. And basing on these dynamical laws for variable  elementary energies, they compute accompanying transport of elementary objects with constant masses $m$ or fixed rest-energies $mc^2$.     These competing force and energy approaches lead to the same low speed dynamics of probe masses $m$ in external fields but not of  material densities $\mu$ within continuous media with inhomogeneous \lq life forces\rq{} $\mu c^2 \beta^2/2$. Streams of liquids and gases are  available for measurements and new conceptual tests to distinguish Newton and Cartesian mechanics for reality.

     The Cartesian acceleration law (28) for mass-energy flows within the Einstein-Infeld material space  can be considered in the limit of vanishing gravitational fields, $g_{\mu\nu} \rightarrow \eta_{\mu\nu} \equiv \{1, -1,-1,-1\}$, and low speeds, $\beta^2 \rightarrow 0$, $B \rightarrow 1- \beta^2/2$, $v^i\rightarrow c dx^i/dx^o \equiv V^i =V_i$, $dV^i/dt =\partial_t V^i +  V^j \partial_j V^i =\partial_t V^i + {\delta^{ij} \partial_j }V^2/2  - [{\bf V}\times curl{\bf V} ]^i$:
          \begin {equation}
\mu ({\partial_t  V^i} + V^j \partial_j V^i)  +   {V^i \partial_t \mu} + \frac {\mu \partial_i V^2}{2} =   -  { \partial_i p }   +\mu \nu   
 \partial_j \partial^j V^i  + f^i_{ext}.
        \end {equation}      
  Here the post Euler  force density $V^i\partial_t \mu$ is the 1903  
  Tsiolkovsky force for the reactive rocket motion.
  The new kinematic feedback  $\mu \partial_i V^2/2$  originates from local gradients of the scalar inertial field (20), which is not relevant to the 1916 GR dynamics in Newtonian empty space. The new kinematic potential $V^2/2$, which modifies Euler's fluid dynamics and the Navier - Stokes equation, appears in (29)   
exclusively in Cartesian physics due to the nondual tensor equation  (26) for the nonempty inertial space. 
   The right-hand side of the equation (29) is associated with all non-relativistic forces from the stress tensor   $P_{\mu}^\nu$ in (28). Here we underline  contributions from the pressure $p$, from the regular drag force with the Stokes kinematic viscosity $\nu$, and from various external forces $f^i_{ext}$, including potential ones.
 
The pressure $p = \mu (sT + \mu_{ch} - \varepsilon) $ within fluids can be related quantitatively, as is  known \cite {Pat, LaLi}, to the temperature $T$, the specific entropy $s$, the chemical potential $\mu_{ch}$, and the specific internal energy $\varepsilon$, with $d\varepsilon = Tds - pd(1/\mu)$ and $d\mu_{ch} = (dp/\mu) -sdT$.   The pressure differential, $dp = (sT + \mu_{ch} - \varepsilon)d\mu + \mu (sdT + d\mu_{ch} +pd \mu^{-1})$ = $\mu (s dT + d\mu_{ch})$, is related only to changes of the local temperature and the chemical potential. The net dynamical balance, 
\begin {equation}
(f_i/\mu) + \nu   
 \partial_j \partial^j V_i- s \partial_i T - \partial_i (\mu_{ch} +  V^2/2) - V_i \partial_t ln \mu   =  dV_i/dt,
 \end {equation}
  of  the Euler material derivative $dV_i/dt$ in the equation (29) depends on speed gradients next to thermal ones, $- [\partial_i (V^2/2) + s\partial_i T]$, along and across material streams.
Such a modification can suggest a lot of conceptual tests to compare Newtonian and Cartesian physics even for steady flows of liquids and gases with $dV_i/dt =0$. For example, contrary to Newtonian physics for laminar fluids with homogeneous pressure and temperature across the radial section of tubes,  Cartesian physics predicts inhomogeneous $p(r)$ and $T(r)$ across the same steady flows. This can shed new light on  \lq unexpected\rq{} heat and pressure on boundaries of many dynamical systems like rocket engines, Ranque-Hilsch vortex tubes \cite{RH}, plasma beams, etc. There are also various applications of the modified Navier-Stokes equation (29) to turbulent flows of energy in industrial hydro generators of electric power.

 The kinematic self-deceleration $-\partial_iV^2/2$ due to the feedback of Leibniz - de Coriolis \lq living forces\rq{} of a moving material space can be understood only through the Cartesian  approach to physical reality.
This inertial self-deceleration prevents asymptotic flows with diverging energy densities.  
The unphysical energy divergence in the original Euler and Navier-Stokes equations was questioned by many researchers, including experts of the Clay Mathematical Institute  (http://www.claymath.org/millennium-problems/navier-stokes-equation).   Newtonian based mechanics  results in the incomplete description of  Navier-Stokes flows, while the Cartesian replica (29) with the inertial damping by \lq living forces\rq{} is a more realistic  approximation of real fluid dynamics.

In general, the new inertial force with its kinematic potential (20) for moving material spaces can justify the nondual field reality of Einstein and Infeld for the macroscopic Cartesian world. Nondual extended matter  refutes applications of Newtonian dynamics to  continuous media and considers the Navier - Stokes  equation with diverging asymptotic solutions as conceptually insolvent. By closing, Cartesian mechanics does deserve very careful investigation for low speed energy flows and for further technological developments  in a line of advanced Einstein's ideas.

\bigskip
\noindent {\bf Discussion}

Experiment oriented  motto  \lq Nullius in verba\rq{}  has been very useful for rapid development of practical applications of science, but eventually neglected the  logical infer  of aether by the Ancient Greeks. The medieval scholastic philosophers of the twelfth to sixteenth centuries assigned inhomogeneous aether of high densities to formation of all visible bodies and the observed nebular planets.  So far there is no direct experimental evidence in favor (or against) of non-empty space. Therefore, any attempts to criticize the empty space paradigm for gravitation  are still considered by the Standard Model to be speculations.  Modern materialistic science with its empty space on the macroscopic scales  is very far away not only from the Eastern philosophies, but even from the classical German Idealism, the Cartesian Natural Philosophy, the Russian Cosmism, and other teachings, which have no specific experimental confirmations. In contrast to the nondual field-energy physics with only one philosophical category for the material reality, the dual physics of  particles and their remote fields is constantly discussing the need or lack of need of new categories, like aether, extra dimensions of the unphysical empty space, etc. 

The nondual Einstein-Infeld physics identified any invisible matter, like aether, with mass-energy density jf the nonempty space with the inhomogeneous Ricci scalar. 
The main concentrations of energy within the infinite radial electron  will always  have spatial dimensions that are much less than the instrumental record of measured space inhomogeneity  (now $10^{-19}m$), because half of the electron charge belongs to the sphere of extremely small radius $r_o \equiv |{\sqrt G}m_o - ie_o|\varphi_o = 1.38 \times 10^{-36}m $.  The other half of the elementary (astro)charge is distributed over micro, macro and mega scales, but corresponding classical densities  probably should not be measured at all. The extended electron  will always be found in the laboratory as an energy point under inelastic exchanges. So what? If one is unable to 
measure fine material densities, should we then claim the emptiness of space  contrary to the logical proof of Aristotle?

It is difficult to understand why contemporary relativists are still working with gravitation of  point masses in question  and with corresponding metric singularities. Why are gravitation experts still calling Einstein  \lq a reluctant father of black wholes\rq{} but not as a science opponent of unphysical metric holes? Einstein has rejected Schwarzschild singularities in 1939 with reasonable scientific arguments \cite {BH}. Also, he very clearly explained together with Infeld that \cite {Ein}: \lq\lq
From the relativity theory we know that matter represents vast stores of energy and that energy represents matter. We cannot, in this way, distinguish qualitatively between matter and field, since the distinction between mass and energy is not a qualitative one. By far the greatest part of energy is concentrated in matter; but the field surrounding the particle also represents energy, though in an incomparably smaller quantity. 	We could therefore say: Matter is where the concentration of energy is great, field where the concentration of energy is small. But if this is the case, then the difference between matter and field is a quantitative rather than a qualitative one. 
There is no sense in regarding matter and field as two qualities quite different from each other. We cannot imagine a definite surface separating distinctly field and matter.
But the division into matter and field is, after the recognition of the equivalence of mass and energy, something artificial and not clearly defined. Could we not reject the concept of matter and build a pure field physics? What impresses our senses as matter is really a great concentration of energy into a comparatively small space\rq\rq{} and 
 \lq\lq We could regard matter as the regions in space where the field is extremely strong. In this way a new philosophical background could be created. Its final aim would be the explanation of all events in nature by structure laws valid always and everywhere. A thrown stone is, from this point of  view, a changing field, where the states of greatest field intensity travel through space with the velocity of the stone. There would be no place, in our new physics, for both field and matter, as fields would be the only reality. This new view is suggested by the great achievements of field physics, by our success in expressing the laws of electricity, magnetism, gravitation in the form of structure laws, and finally by the equivalence of mass and energy. Our ultimate problem would be to modify our field laws in such a way that they would not break down for regions in which the energy is enormously concentrated.\rq\rq{}
Why was the pure field physics of Einstein silently ignored by \lq Nullius in verba\rq{}  science? How may one speculate whether Einstein accepted  or not excepted aether as a third notion if he publicly suggested to transform massive particles into fields without any extra notions?

Einstein's energetic way for unification of the continuous mass and its gravitational field works equally well to describe the continuous electric charge in terms of Maxwell classical fields.
Again, Maxwell-Mie electron\rq{s} density scale, $e_o{\sqrt G}/ c^2 = 1,38\times 10^{-36}m $ is even  less than the Planck's length, $l_p \equiv {\sqrt {\hbar c}} ({\sqrt G}/ c^2) = 1,62 \times 10^{-35}m$. Such scales for elementary matter concentration can explain the success of the $\delta$-operator modeling of the nonlocal energy carrier.  However, formal interactions of point-like concentrations of charge with  external electric fields is conceptually incorrect for real continuous particles because of their direct and global superimposition in the nonlocal Universe.  Despite the obvious observation of empty space, in practice, Dirac delta density should be used only for engineering calculations, and not for conceptual approaches to reality (non-empty space, yin-yang pairing) of continuously charged fields.    
Indeed, the well-performed measurements of electric fields associated with moving  charges \lq\lq{}support the idea of a Coulomb field carried {\it rigidly} by the electron beam\rq\rq{} \cite{Piz, Pizz}. These experiments confirm directly the nonempty space solution (3) where the elementary charge is continuously distributed over the radial Coulomb field.   Similar nonempty space solutions can be derived in General Relativity for the extended mass \cite {Bul}. 

By referring to the yin-yang dialectic of paired energy flows with time-varying Coulomb fields in the EM experiments \cite{Piz, Pizz}, the nonempty space concept can equally predict the local coupling of inertial and gravitational energy flows. These nonempty space options with low-frequency sound waves within the gravitational mass-energy continuum should be counted in LIGO and VIGRO data processions. Like in wave electrodynamics, mechanical waves in the continuous  material medium do follow cosmological events. So far there is no concrete evidence to claim that gravitational/inertial signals have to be considered as a confirmation of the event horizon. The latter backs the black hole in the Newton dual model of reality.

In fact, Einstein's gravitation was tested only in the weak field experiments. The same model of Newtonian point-like particles can quantitatively describe both classical dynamics and its post-Newtonian corrections of the first order in empty (but curved) space. Such tests may only state that Newton and Schwarzschild were right for weak field approximations. At the same time, 
nonempty (but flat) space physics provides \cite {BulR}   the measured post-Newtonian corrections in the main gravitational tests.  
Actually, Descartes and Einstein-Infeld were right for both weak and strong material fields. The simplified Newtonian interpretation of reality through distant material peculiarities leads to the vulgarization of Einstein's gravitation. Cartesian physics is free from peculiarities and the black hole notion. Quantum mechanics has already justified the extension of elementary matter over all spatial scales. Today the turning point for each researcher is whether to follow Einstein's, Descartes's, and Aristotle's descriptions of the non-empty space mechanics and gravitation,  or to remain adherent to the empty space paradigm of Newton.

The author is confident that the Cartesian world alternative in testable mathematical predictions is significant for all areas of natural sciences, both fundamental and applied.  Contrary to the empty Newtonian space with localized matter, like in the contemporary textbooks on mechanics and gravitation, the nonempty Cartesian space of continuously distributed matter has real chances for the ontological convergence of classical and quantum physics in the near future. Various applications of material space begin with the Earth sciences, biology, medicine, etc., and ends with astrophysics and the continuous microcosm of extending astroparticles.

The radial distribution of the elementary carrier of mechanical and electrical complex energies (4) is a first step toward unification of nonlocal classical and quantum matter. The half-charge scale $r_o \equiv e_o/\varphi_o =1,38 \times 10^{-36}m $ and the minimal radius $r_m = \hbar c/e_o\varphi_o = 1,89 \times 10^{-34}m $ of the 
Sommerfeld quantization loop   can shed   
 new light on the physical origin of the Sommerfeld (fine-structure) constant, $a_o/r_m = 1/137$, and on the physical meaning of the Plank length, $(\hbar c /\varphi_o)^{1/2} \equiv   (r_ma_o)^{1/2} = 1.62\times 10^{-35}m$, directly from the Maxwell theory.
In closing,  the author  encourages  all researchers  to reject particles as the independent concept (in spite of  observations of spatially separated bodies)  and to come back to Mie \cite {Mie} and Einstein \cite {Ein} nondual ideas for better evolution of contemporary physics.  All natural sciences, guided by the Yin-Yang equivalence of active / passive complex charge densities in paired energy flows, can together decipher the ancient Chinese approach to the kinetic energy reality in our non-local universe with heated material space.

\end{document}